\documentclass{article}
\usepackage{amssymb}
\usepackage{amsmath}
\begin{document}
\centerline{\Large \bf Formal exact operator solutions to nonlinear differential equations }
\vskip 2cm
\centerline{\sc Yu. N. Kosovtsov}

\medskip
\centerline{Lviv Radio Engineering Research Institute,79060 Lviv, Ukraine}
\centerline{email: {\tt yunkosovtsov@gmail.com}} \vskip 1cm
\begin{abstract}
The compact  explicit expressions for formal exact operator solutions to Cauchy problem for sufficiently general systems of nonlinear differential equations (ODEs and PDEs) in the form of chronological operator exponents are given. The variant of exact solutions in the form of ordinary (without chronologization) operator exponents are proposed.
\end{abstract}

\section{Basis of the method.} The starting point for solving differential equations by the operator method  is \emph{linear} first-order differential equation for operator  ${\bf E}={\bf E}(t)$:
\begin{equation}
\frac{\partial {\bf E}}{\partial t} ={\bf L}(t){\bf E}
\label{dE}
\end{equation}
with the initial condition
\begin{equation}
{\bf E}|_{t=a} = {\bf I},
\label{dEin}
\end{equation}
where ${\bf L}(t)$ is a \emph{linear operator} which, generally speaking, depends on $t$, but
does not depend on operator $\frac{\partial}{\partial t}$ explicitly; ${\bf I}$ is identity operator.
The solution of this equation for $t>a$ was obtained in 1949 by F. Dyson in the chronological
exponential form \cite {Dyson} (see also \cite {Bogolubov}, \cite {Kirzhnits})
\begin{equation}
{\bf E}(t) ={\bf T}\exp \{\int_a^t d\tau\,{\bf L}(\tau)\},
\label{ET}
\end{equation}
where it is supposed, that exponent represents a power series expansion with following \emph{chronologization} according to the rule

\begin{align}
{\bf T}\, \{{\bf L}(\tau_1)\,\ {\bf L}(\tau_2)\dots {\bf L}(\tau_n) \}=&{\bf L}(\tau_{\alpha_1})\,\ {\bf L}(\tau_{\alpha_2})\dots {\bf L}(\tau_{\alpha_n}) . \notag\\ &
 \tau_{\alpha_1}\geq \tau_{\alpha_2}\geq \dots \geq \tau_{\alpha_n}\notag
\end{align}

Very important role plays the equation adjoint to (\ref{dE}) for operator ${\bf E}^{-1}={\bf E}^{-1}(t)$ (see \cite {Gam1}, \cite {Gam2})
\begin{equation}
\frac{\partial {\bf E}^{-1}}{\partial t} =-{\bf E}^{-1}{\bf L}(t)
\label{dE1}
\end{equation}
with the initial condition
\begin{equation}
{\bf E}^{-1}|_{t=a} = {\bf I}. \label{dEin1}
\end{equation}
It is easy to see using the Dyson's idea, that its solution is
\begin{equation}
{\bf E}^{-1}(t)={\bf T}_0\exp \{-\int_a^t d\tau\,{\bf L}(\tau)\},
\label{T00}
\end{equation}
where operator ${\bf T}_0$ is the operator of opposed chronologization with action
\begin{align}
{\bf T}_0\, \{{\bf L}(\tau_1)\,\ {\bf L}(\tau_2)\dots {\bf L}(\tau_n) \}={\bf L}&(\tau_{\alpha_1})\,\ {\bf L}(\tau_{\alpha_2})\dots {\bf L}(\tau_{\alpha_n}) . \notag\\ & \tau_{\alpha_1}\leq \tau_{\alpha_2}\leq \dots \leq \tau_{\alpha_n}\notag
\end{align}

It appears that the operator ${\bf E}^{-1}(t)$ is inverse to the ${\bf E}(t)$, i.e.
\[
{\bf T}_0\exp \{-\int_a^t d\tau\,{\bf L}(\tau)\}\,{\bf T}\exp \{\int_a^t d\tau\,{\bf L}(\tau)\}={\bf I}
\]

Ordering operators ${\bf T}$ and ${\bf T}_0$ for $t<a$ change places. In some sense in linear problems operators ${\bf T}$ and ${\bf T}_0$ provide description of direct and back scattering accordingly. It be assumed henceforth that $t>a$.

Ordering (chronological) operators have to be introduced because the operators in the
integrands with different values of the variable $\tau$ may not commute, what cause certain difficulties in manipulation with chronological exponents. Nevertheless, examination of algebraical properties of the chronological exponents allows to obtain an extensive family
of operator identities \cite {Gam1}-\cite {Kos1} which provide possibilities of transformations and analysis of chronological expressions.

The operator ${\bf \Delta}$ is \emph {derivative} if
\[{\bf \Delta}({\bf A + B })f = {\bf \Delta A} f + {\bf \Delta B} f,\]
\[{\bf \Delta}{\bf A}f = ({\bf \Delta A}) f + {\bf A}{\bf \Delta} f\]
for any differentiable function $f$ and  any differentiable linear operators ${\bf A}$ and ${\bf B}$. In other words, the linear operation is derivative if it satisfies to Leibnitz rule. As we will see later, the chronological operators with derivative operators in exponent, which we will denote for short as
\[
{\bf E} ={\bf T}\exp \{\int_a^t d\tau\,{\bf \Delta}(\tau)\},
\]
 play the extremely important role here.

Let $b(t)$ be a \emph{function} that is the operator of
multiplication on the $b(t)$. Then, as can be seen from the definition of
the derivative operator, the commutator $[b(t),{\bf \Delta}(\tau)]$
is a function too.

Now consider the following construction
\[
{\bf K}={\bf E}\,b(t)\,{\bf E}^{-1}.
\]
As far as the commutator $[b(t),{\bf \Delta}(\tau)]$ is a function,
then all repeated commutators are functions too. So with help of BCH
formula (see (27) in \cite {Kos1}) we conclude that operator ${\bf K}$ is \emph
{a function} and
\begin{equation}
{\bf K}=({\bf E}\,b(t)), \label{hom}
\end{equation}
where the outward brackets in the right-hand side stress that action of operator is restricted by the brackets area.
An analogous conclusion holds for ${\bf E}^{-1}\,b(t)\,{\bf E}$.

Let us consider further the following obvious chain
\begin{align}
{\bf E}\,b_1\,b_2\,\dots\,b_n&={\bf E}\,b_1\,{\bf E}^{-1}\,{\bf
E}\,b_2\,{\bf E}^{-1}\,\dots\,{\bf E}\,b_n= \notag
\\&=({\bf E}\,b_1)\,({\bf E}\,b_2)\,\dots\,(\,{\bf E}\,b_n). \notag
\end{align}

With what is observed here and the fact that the chronological operator is linear,
leads to conclusion that for any function
$F(b_1,\,b_2,\,\dots,\,b_n)$, which can be expanded in power series
with respect to $b_1,\,b_2,\,\dots,\,b_n$, we can obtain the
following nice property:
\begin{equation}
{\bf E}\,\,F(b_1,\,b_2,\,\dots,\,b_n)=F(({\bf E}\,b_1),\,({\bf
E}\,b_2),\,\dots,\,({\bf E}\,b_n)). \label{homom}
\end{equation}

Analogously
\begin{equation}
{\bf E}^{-1}\,\,F(b_1,\,b_2,\,\dots,\,b_n)=F(({\bf
E}^{-1}\,b_1),\,({\bf E}^{-1}\,b_2),\,\dots,\,({\bf E}^{-1}\,b_n)).
\label{homom0}
\end{equation}

This property of the operator ${\bf E}$ coincides with familiar property of the shift operator $\exp\{s\frac{\partial}{\partial b}\}$, as shift operator is the particular case of chronological exponent.

It be assumed henceforth that $F_i$  are analytic functions on all their arguments in corresponding regions.

\section{Chronological solutions of nonlinear differential equations.}

In existent literature, devoted to operator approach for solving differential equations, as a rule, the linear problems are considered, i.e., one proceed from  (\ref{dE})-(\ref{ET}). The connection of equations (\ref{dE}) and (\ref{dE1}) with nonlinear \emph{ordinary} differential equations seemingly for the first time was explicitly mentioned in \cite {Gam1}, \cite {Gam2}. The similar attempt to nonlinear \emph{partial} differential equations in \cite {Agr} had not led to compact form of solution, only infinite chronological iteration  series was obtained there.

Let us consider particular cases of the derivative operators ${\bf \Delta}$ and differential equations, which correspond them.

\subsection {${\bf \Delta_1}(\tau)=F(\tau,c)\frac{\partial}{\partial c}.$}
Differentiating function
\begin{equation}
u(t,c) ={\bf T}_0\exp \{-\int_a^t d\tau\,F(\tau,c)\frac{\partial}{\partial c}\}\,c
\label{odesol}
\end{equation}
on $t$, we obtain, taking into account (\ref{dE1}) and then (\ref{homom0}), that
\[
\frac{\partial u}{\partial t} ={\bf T}_0\exp \{-\int_a^t d\tau\,F(\tau,c)\frac{\partial}{\partial c}\}F(t,c)=-F(t,u),
\]
i.e, the expression (\ref{odesol}) is a solution of ordinary differential equation of the \emph{first order} for $u=u(t,c)$ (see \cite {Gam1}, \cite {Gam2}, and also \cite {Kos1})
\begin{equation}
\frac{\partial u}{\partial t} +F(t,u)=0,
\label{ode}
\end{equation}
where $c$ is an arbitrary constant, $u|_{t=a} = c$. That is, the expression (\ref{odesol}) is exact \emph{general} operator solution of the equation (\ref{ode}).

\subsection {${\bf \Delta_2}(\tau)=\sum_{j=1}^n F_j(\tau,c_1,\dots,c_n)\frac{\partial}{\partial c_j}.$}
Analogously, differentiating functions $u_i=u_i(t,c_1,\dots,c_n)$, where $i=1,\dots,n$,
\begin{equation}
u_i ={\bf T}_0\exp \{-\int_a^t d\tau\,\sum_{j=1}^n F_j(\tau,c_1,\dots,c_n)\frac{\partial}{\partial c_j}\}\,c_i
\label{sysodesol}
\end{equation}
on $t$, we ascertain that  (\ref{sysodesol}) (see \cite {Kos1}) is the solution of the following \emph{system} of ordinary differential equations of the \emph{first order} for $u_i=u_i(t,c_1,\dots,c_n)$
\begin{equation}
\frac{\partial u_i}{\partial t} +F_i(t,u_1,\dots,u_n)=0,\qquad (i=1,\dots,n)
\label{sysode}
\end{equation}
where $c_i$ are arbitrary constants, $u_i|_{t=a} = c_i$.

\subsection {${\bf \Delta_3}(\tau)=\int_{-\infty}^\infty d^m \zeta \,\sum_{j=1}^n F_j(\tau,\vec{\zeta},\vec{c},\dots,D_\zeta^\alpha c_k,\dots)\frac{\delta}{\delta c_j(\vec{\zeta})},$}
where $\tau$ is dedicated independent variable ("time" variable),  $\vec{\zeta}=\zeta_1,\dots,\zeta_m$ are "space" variables, the functions $\vec{c}=\vec{c}(\vec{\zeta})=c_1(\vec{\zeta}),\dots,c_n(\vec{\zeta})$  will be later on play role of arbitrary functions of "space" variables (it is clear, that it is supposed that all these functions are arbitrary smooth and have not singularities), the symbol  $\dots,D_\zeta^\alpha c_k,\dots$ denotes a set of all partial derivatives of $c_k$ ($k=1,\dots,n$) on "space" variables $\zeta_i$ of the type \[D_\zeta^\alpha c_k=\frac{\partial^{\alpha_1}\dots\partial^{\alpha_m}c_k}{\partial \zeta_1^{\alpha_1}\dots\partial \zeta_m^{\alpha_m}}\]
up to some certain order,  $\frac{\delta}{\delta c_j(\vec{\zeta})}$ is the functional derivative.

Differentiating on $t$ functions
\begin{equation}
u_i(t,\vec{x}) ={\bf T}_0\exp \{-\int_a^t d\tau\, {\bf \Delta_3} \}\,c_i(\vec{x})\,,
\label{syspdesol}
\end{equation}
where $\vec{x}=x_1,\dots,x_m$, we obtain, taking into account (\ref{dE1}) and then (\ref{homom0}), that
\begin{align}
&\frac{\partial u_i}{\partial t}=-{\bf T}_0\exp \{-\int_a^t d\tau\, {\bf \Delta_3} \}\,\circ\notag
\\& \int_{-\infty}^\infty d^m \zeta \,\delta(\vec{\zeta}-\vec{x})\,F_i(t,\vec{\zeta},\vec{c},\dots,D_\zeta^\alpha c_k,\dots)=
\notag\\&=-{\bf T}_0\exp \{-\int_a^t d\tau\, {\bf \Delta_3} \}\,F_i(t,\vec{x},\vec{c},\dots,D_x^\alpha c_k,\dots)=
\notag\\&=-F_i(t,\vec{x},\vec{u},\dots,D_x^\alpha u_k,\dots)\notag
\end{align}
and ascertain that (\ref{syspdesol}) is the solution of  \emph{system} of \emph{partial} differential equations of the \emph{first order} on variable $t$ (and arbitrary order on "space" variables) for $u_i(t,\vec{x})$
\begin{equation}
\frac{\partial u_i}{\partial t} +F_i(t,\vec{x},\vec{u},\dots,D_x^\alpha u_k,\dots)=0,\qquad (i=1,\dots,n)
\label{syspde}
\end{equation}
where $\vec{u}=u_1(t,\vec{x}),\dots,u_n(t,\vec{x})$ and now in $\dots,D_x^\alpha u_k,\dots$ appears the corresponding set of partial derivatives of $u_k(t,\vec{x})$ on "space" variables $\vec{x}$.

For equations of higher orders on $t$ by substitutions of type $\frac{\partial u_i}{\partial t}=v_i$ very often one can get over to system of the first order equations and, if it possible to resolve such system with respect to derivatives of unknown functions (including $v_i$) on $t$ (i.e., reducing the equation or system into the so-called normal form), it is easy enough by using 2.2 or 2.3 to obtain operator solutions for these cases too.

The existence of formal solutions of nonlinear differential equations in the form of chronological operator exponents is important as a practical matter, first of all, because they represent in compact and clear form the dependence of solutions from all parameters of the problem. If given differential equation fits the conditions on existence and uniqueness of solutions, then all forms of exact solutions are equivalent (with minor reservations) with each other. However, in usage of one or another form of solution one is wishing to have the opportunity to analyse it and certain freedom in its transformation in follow-on calculations. The interpretation of chronological solutions as infinite series is the first, but far from being an only opportunity. There are simple examples, when operator solution can be transformed to the conventional form just by operator methods (see, e.g., \cite {Kos1}).

Fortunately for the stage of analytic (symbolic) calculations there is an extensive family of operator identities \cite {Gam1}-\cite {Kos1}, which allow to carry out not only basic algebraic operations with chronological exponents, but accomplish its differentiation on parameters and so on. If you wish, algebraic transformations here can be interpreted as resummation of corresponding infinite iteration series in which compactness, visibility are kept and, what is important, opportunity to remain as soon as possible in terms of exact solutions.

Nevertheless on the stage of the decision-result ("number") obtaining very often it is necessary to turn to series. The presence of chronologization, which is overburdened by multiple integration, complicates considerably as calculations as essential analysis of series convergence conditions. Therefore during more than half a century the attempts to represent operator solutions in form of ordinary (without chronological operator) exponent are made (see survey \cite {Oteo}), here one of typical approaches is the so-called Magnus expansion. There are some variants to solve this problem, which have its own benefits depending on posed a goal. As a rule such approaches lead to approximate presentation in the form of very bulky expressions. The variant proposed below is differed by noticeable simpleness and is, what is more, the exact solution.

\section{The solutions of nonlinear differential equations in form of ordinary (without chronological operator) exponent.}

It is follows from operator identity (23) of \cite {Kos1} and well-known properties of shift operator that (note, operator {\bf L}(t) does not depend on operator $\frac{\partial}{\partial t}$ explicitly)

\begin{align}
&\exp\{(t-a)\frac{\partial}{\partial s}\}\exp\{(t-a)[{\bf L}(s)-\frac{\partial}{\partial s}]\}=\notag\\\notag
\\&={\bf T}\exp \{\int_a^t d\tau\,{\bf L}(\tau+s-a)\}\,,\notag
\end{align}
from here
\begin{align}
&\exp\{(t-a)[{\bf L}(s)-\frac{\partial}{\partial s}]\}=
\notag\\
\notag\\=&\exp\{-(t-a)\frac{\partial}{\partial s}\}{\bf T}\exp \{\int_a^t d\tau\,{\bf L}(\tau+s-a)\}=
\notag\\
\notag\\=&{\bf T}\exp \{\int_a^t d\tau\,{\bf L}(\tau+s-t)\}\exp\{-(t-a)\frac{\partial}{\partial s}\}\,.\notag
\end{align}

or equivalent
\begin{eqnarray}
\exp\{(t-a)[{\bf L}(s)-\frac{\partial}{\partial s}]\}\exp\{(t-a)\frac{\partial}{\partial s}\}=
\nonumber\\
={\bf T}\exp \{\int_a^t d\tau\,{\bf L}(\tau+s-t)\}\,.\notag
\end{eqnarray}
Similarly for inverse operators

\begin{eqnarray}
\exp\{-(t-a)[{\bf L}(s)-\frac{\partial}{\partial s}]\}\exp\{-(t-a)\frac{\partial}{\partial s}\}=\nonumber
\\={\bf T}_0\exp \{-\int_a^t d\tau\,{\bf L}(\tau+s-a)\}\,.\label{exp0}
\end{eqnarray}

Let us apply the obtained identities to the solutions of differential equations of Section 2. Note, first of all, that if both sides of identity (\ref{exp0}) multiply from the right on constant $c_k$ or function $c_k(\vec{x})$, which do not depend on $s$, then the action of right shift operator in the left-hand side will be equal to the action of identity operator $I$, while left-hand side operator under \emph{substitution} $s=a$ turns into above examined operator (\ref{T00}).

Thus for
\[{\bf L}(\tau)={\bf \Delta_1}(\tau)=F(\tau,c)\frac{\partial}{\partial c}\]
we have, taking into account (\ref{exp0}) and foregoing remarks, that
\begin{eqnarray}
u(t) =[{\bf T}_0\exp \{-\int_a^t d\tau\,F(\tau+s-a,c)\frac{\partial}{\partial c}\}\,c]|_{s=a}=\nonumber
\\=\exp\{-(t-a)[F(s,c)\frac{\partial}{\partial c}-\frac{\partial}{\partial s}]\}\,c\,|_{s=a}.
\nonumber
\end{eqnarray}
As a result we get the solution of equation (\ref{ode}) in exponential form (without chronological operator)
\[
u(t) =\exp\{-(t-a)[F(s,c)\frac{\partial}{\partial c}-\frac{\partial}{\partial s}]\}\,c\,|_{s=a}.
\]

By the same way one can obtain the exponential form of solution for system of ordinary differential equations (\ref{sysode})
\[
u_i(t) =\exp\{-(t-a)[\sum_{j=1}^n F_j(\tau,c_1,\dots,c_n)\frac{\partial}{\partial c_j}-\frac{\partial}{\partial s}]\}\,c_i\,|_{s=a} , \qquad (i=1,\dots,n).
\]

Similarly the exponential form of solution for system of partial differential equations (\ref{syspde}) is as follows
\[
u_i(t,\vec{x})  =\exp\{-(t-a)[{\bf \Delta_3}(s)-\frac{\partial}{\partial s}]\}\,c_i(\vec{x})\,\mid_{s=a} , \qquad (i=1,\dots,n).
\]

If expand the operator exponent into series, then one can obtain the expansion of the solution into Taylor series on powers of $(t-a)$. This is most obvious (but not unique) interpretation of the operator exponential form of solutions.

The \emph{Maple} procedures for analytical calculations of approximate solutions of ordinary and partial differential equations (Cauchy problem), which are based on the approach considered here, are presented in \cite {Kos2}.


\begin{thebibliography}{9}

\bibitem {Dyson} Dyson F J 1949 {\it Phys. Rev.} {\bf 75} 486

\bibitem {Bogolubov} Bogoliubov N N and Shirkov D V 1983 {\it Quantum fields} (New York: Benjamin-Cummings) p~388

\bibitem {Kirzhnits} Kirzhnits D A 1967 {\it Field Theoretical Methods in Many-Body Systems} (Oxford: Pergamon) p~410

\bibitem {Gam1} Agrachev A A and Gamkrelidze R V 1978 {\it Matem. sbornik} {\bf 107} 467

\bibitem {Gam2} Agrachev A A and Gamkrelidze R V 1980 {\it Itogi nauki. VINITI. Problemy geometrii} {\bf 11} 135

\bibitem {Kos1} Kosovtsov Yu N 2004 The Chronological Operator Algebra and Formal Solutions of Differential Equations {\it Preprint} math-ph/0409035

\bibitem {Agr} Agrachev A A and  Vakhrameev S A 1981 {\it Itogi nauki. VINITI. Problemy geometrii} {\bf 12} 165

\bibitem {Oteo}Blanes S, Casas F, Oteo J A, Ros J 2009 {\it Physics Reports} {\bf 470} n. 5-6 151

\bibitem {Kos2} Yu.\,N. Kosovtsov, http://www.maplesoft.com/applications/view.aspx?SID=1690;\\ http://www.maplesoft.com/applications/view.aspx?SID=1687; \\ http://www.maplesoft.com/applications/view.aspx?SID=1426


\end{thebibliography}
\end{document}